\documentclass{article}
\usepackage{spconf,amsmath,graphicx}
\usepackage{makecell}
\usepackage{amsfonts}

\usepackage{flushend}
\usepackage{algorithm}
\usepackage{algorithmicx}
\usepackage{mathtools}
\usepackage{algpseudocode}
\usepackage{subfigure}
\usepackage{multirow}
\usepackage{longtable}
\usepackage{amsthm}
\usepackage{amssymb }
\usepackage{enumitem}

\usepackage{bm}
\usepackage{mathrsfs}

\usepackage{longtable}
\usepackage{multirow}
\usepackage{cite}
\usepackage{xcolor}
\usepackage{graphicx}

\def\KMM{\text{MAST}}

\usepackage{soul}
\title{Masked Autoencoder with Swin Transformer Network for Mitigating Electrode Shift in HD-EMG-based Gesture Recognition}

\name{Kasra Laamerad$^\dagger$, Mehran Shabanpour$^\dagger$, Md. Rabiul Islam$^\ddagger$, Arash Mohammadi$^\dagger$}
\address{$~^\dagger$Concordia Institute for Information Systems Engineering, Concordia University, Montreal, Canada\\
$~^\ddagger$ Nanyang Technological University (NTU), Singapore}

\begin{document}
\ninept
\maketitle
%
%============================================================
\begin{abstract}
Multi-channel surface Electromyography (sEMG), also referred to as high-density sEMG (HD-sEMG), plays a crucial role in improving gesture recognition performance for myoelectric control. Pattern recognition models developed based on HD-sEMG, however, are vulnerable to changing recording conditions (e.g., signal variability due to electrode shift). This has resulted in significant degradation in performance across subjects, and  sessions.  In this context, the paper proposes the Masked Autoencoder with Swin Transformer ($\KMM$) framework, where training is performed on a masked subset of  HD-sEMG channels.  A combination of four masking strategies, i.e., random block masking; temporal masking; sensor-wise random masking, and; multi-scale masking, is used  to learn latent representations and increase robustness against electrode shift. The masked data is then passed through $\KMM$'s three-path encoder-decoder structure, leveraging a multi-path Swin-Unet architecture that simultaneously captures time-domain, frequency-domain, and magnitude-based features of the underlying HD-sEMG signal.  These augmented inputs are then used in a self-supervised pre-training fashion to improve the model's generalization capabilities. Experimental results demonstrate the superior performance of the proposed $\KMM$ framework in comparison to its counterparts. 
\end{abstract}
%============================================================
%
%============================================================
%============================================================
%
%%%%%%%%%%%%%%%%%%%%%%%%%%%%%%%%%%%%%%%%%%%%%%
\section{Introduction} \label{sec:intro}
%%%%%%%%%%%%%%%%%%%%%%%%%%%%%%%%%%%%%%%%%%%%%%

Electromyography (EMG) is widely used for recording muscle activities, playing a crucial role in Human-Machine Interaction (HMI) systems such as myoelectric control~\cite{Bi-2019-2}. Surface electromyography (sEMG)-based gesture recognition is of particular interest in non-invasive applications, such as controlling prosthetic or exoskeletons~\cite{Farina-2014-4}. In this context, a key challenge arises from signal variability~\cite{Islam:2024, Farokh:2024, Kia:2024, Li-2020-6, Ding-2019-7} such as electrode shifts, which occur during normal use, such as putting on or taking off a prosthesis or due to factors such as sweating. Electrode shifts alter the data distribution, leading to significant drops in system performance, often reducing gesture recognition accuracy by up to $25$\% with even minor shifts~\cite{Hargrove-2006-8}. Addressing distribution shift of non-stationary sEMG signals is, therefore, of significant importance for improving the reliability and effectiveness of sEMG-based control systems in real-world applications. 
Conventional methods to mitigate electrode shifts include recalibration techniques~\cite{Zhu-2016-20}. Such methods, however, involves time-consuming processes to collect labelled data after each shift limiting their practical applicability. An alternative approach is capturing redundant signals via high-density sEMG (HD-sEMG)~\cite{Zhang-2019-13} to enhance robustness, which is the focus of this study. 

\vspace{.025in}
\noindent
\textbf{Related Works:}
Traditional Machine Learning (ML) models, such as Linear Discriminant Analysis (LDA) and Support Vector Machines (SVMs), in combination with time-domain and frequency-domain feature extraction methods~\cite{Li-2020-25} form the foundation of early research on gesture recognition using sEMG. 
Feature extraction have evolved from simple Time-Domain (TD) metrics, such as root-mean-square, to Frequency-Domain (FD) features, such as median frequency, and their combination, to more more advanced techniques such as wavelet transforms and Hilbert-Huang transforms. Traditional methods faced challenges in capturing the complex nature of sEMG signals. Furthermore, redundant features and computational inefficiencies hindered the performance of classical ML classifiers. Finally, such approaches often fall short in real-world applications due to their sensitivity to electrode shift and signal variability, as highlighted in~\cite{Stango-2014-32}. 

Consequently, there has been a paradigm shift towards application Deep Learning (DL) techniques. DL methods~\cite{ElahehTNSRE:2021} emerged as a powerful alternative to conventional ML models, enabling automatic feature extraction and learning. 
Convolutional Neural Networks (CNN)-based architectures~\cite{Rami:2024, Elaheh:G2019, Pereira-2024-62}, typically, outperformed traditional classifiers by handling the complexities of sEMG data more effectively. Performance gaps, however, remain, particularly in maintaining robustness across sessions resulting in development of hybrid architectures combining CNN with Recurrent Neural Network (RNN)~\cite{Zhang-2020-54} or Bidirectional Long Short-Term Memory (BiLSTM) layers~\cite{Karnam-2022-55, Kia:2020}. Despite their success, however, CNN-based models are often computationally expensive and require large amounts of labeled data for training, limiting their scalability in real-time applications~\cite{Sun-2022-56, Qureshi-2023-57}. In addition, multi-scale CNNs have been used to capture information from various signal resolutions~\cite{Han-2021-60, Shen-2020-61}.
Attention mechanisms~\cite{SoheilTNSR2023, MansooreEMBC:2022, Montazerin-2023-14} have also been integrated  to improve gesture recognition by enhancing the model's focus on key features within the sEMG signals. References~\cite{Hao-2021-58, Wang-2022-59} demonstrated that adding attention mechanisms to CNN or CNN-RNN architectures can increase recognition accuracy by highlighting the most relevant temporal and spatial features.  These advancements highlight the shift towards more sophisticated DL models capable of learning directly from sEMG data while addressing the limitations of traditional hand-crafted feature extraction.
Addressing electrode shift in sEMG-based gesture recognition systems involves several key strategies. Transformation-based corrections such as  the Shifts Estimation and Adaptive Correction (SEAR)~\cite{Li-2020-6} method and Electrode Shift Fast Correction (ESFC)~\cite{Wang-2023-60} realign shifted signals using correction matrices or gestures. Data augmentation techniques, such as varied electrode placements~\cite{Pereira-2024-62} and the Array Barrel Shifting Data Augmentation (ABSDA)~\cite{Chamberland-2023-61} method, create synthetic data to train models for improved robustness. Transfer learning~\cite{Islam:2024} allows models to adjust to new sessions and mitigate signal variability. 
%Lastly, HD electrode arrays enhance signal stability by using dense electrode configuration~\cite{Chamberland-2023-63, Wang-2024-64}. 
Effectiveness of these methods is constrained by the need for precise shift estimation and limitations in simulations and data availability.

\vspace{.025in}
\noindent
\textbf{Contributions:}
Capitalizing on the above discussion and following the work of~\cite{Pereira-2024-62}, the paper proposes the Masked Autoencoder with Swin Transformer ($\KMM$) framework, where training is performed on masked subset of  HD-sEMG channels selected in varying fashion. Different from~\cite{Pereira-2024-62} where a pre-defined and fixed subset of HD-sEMG channels is used, the proposed $\KMM$ framework builds upon strength of Masked Autoencoder (MAE)~\cite{He:2022} as an scalable learning mechanism to improve resiliency to electrode shift. The $\KMM$ framework consists of a combination of four masking strategies, i.e., random block masking; temporal masking; sensor-wise random masking, and; multi-scale masking. More specifically, first, we leverage a multi-path Swin-Unet architecture that simultaneously captures time-domain, frequency-domain, and magnitude-based features, providing a comprehensive understanding of the sEMG signals without relying on manual feature extraction. In addition, our use of Partial Convolution and advanced masking techniques enhances the model's robustness to missing or corrupted data, outperforming traditional and CNN-based approaches that struggle with electrode shifts. By incorporating cross-modality attention mechanisms and self-supervised pre training, our method efficiently aligns complementary features from different signal representations, improving generalization across different recording sessions and subjects. Lastly, our model achieves high classification accuracy with lower computational requirements, positioning it as a more scalable and practical solution for real-world prosthetic control and human-computer interaction systems.
In summary, the paper makes the following contributions:
\begin{itemize}
\item We show that random masking of channels of a HD-sEMG system through  MAs results in less sensitivity to electrode shifts compared to the scenario where a pre-defined subset is selected.
\item The proposed $\KMM$ framework achieved considerably higher  intersession accuracy compared to its state-of-the-art counterpart.
\end{itemize}
The reminder of the paper is organized as follows: The proposed $\KMM$ framework is introduced in Section~\ref{sec:method}. Experimental results are presented in Section~\ref{sec:results}. Finally, Section~\ref{sec:conc} concludes the paper.

%OOOOOOOOOOOOOOOOOOOOOOOOOOOOOOOOOOOOOOOOOOOOOOOOOOOOOOO
\section{The Proposed $\KMM$ Framework} \label{sec:method}
%OOOOOOOOOOOOOOOOOOOOOOOOOOOOOOOOOOOOOOOOOOOOOOOOOOOOOOO
%%%%%%%%%%%%%%%%%%%%%%%%%%%%%%%%%%%%%%%%%%%%%%%%%%%
\begin{figure}[t!]
    \centering
    \includegraphics[scale=.2]{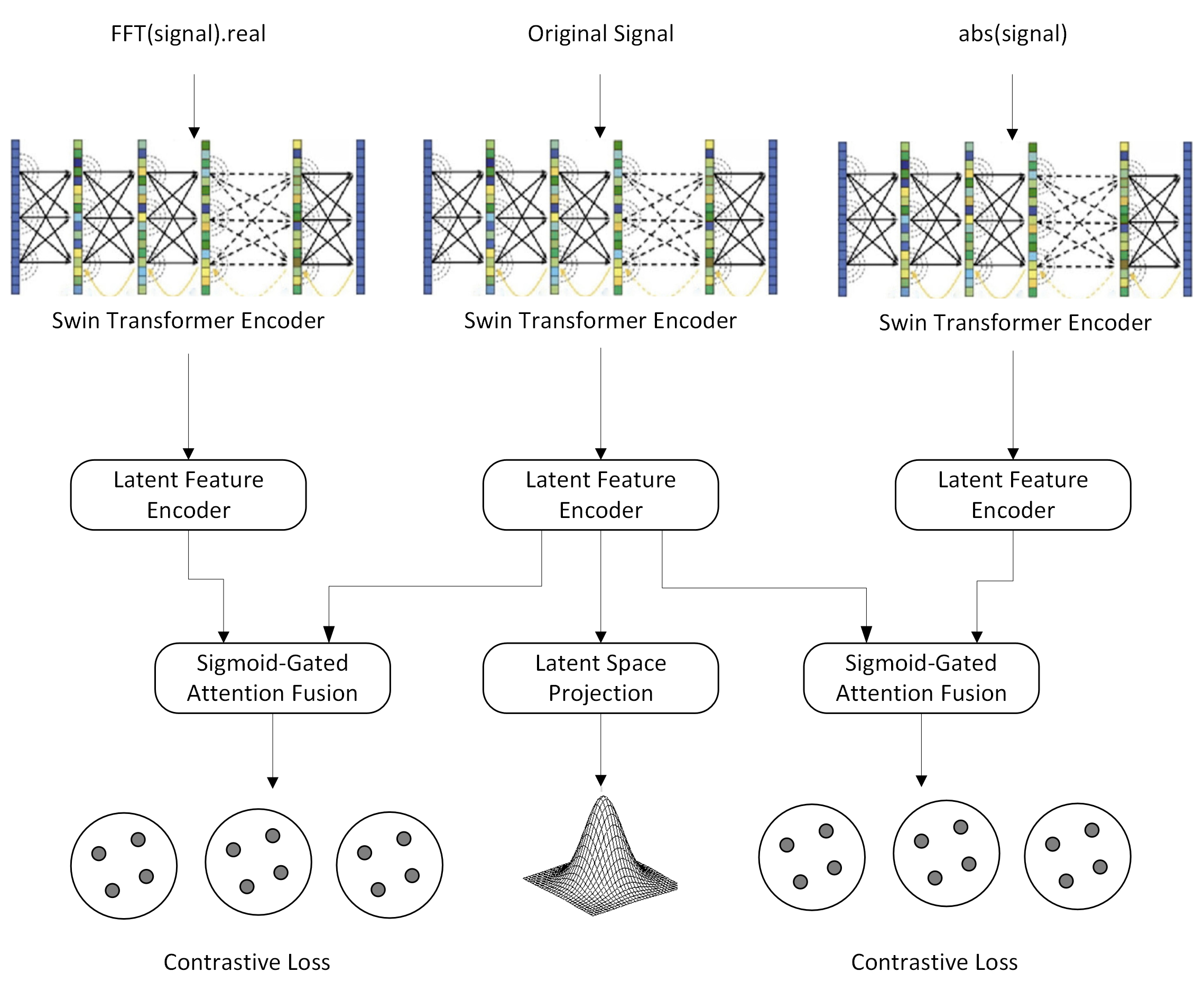}
    \caption{\footnotesize Three-path encoder of the proposed $\KMM$ framework.}
    \label{fig:KMM}
\end{figure}
%%%%%%%%%%%%%%%%%%%%%%%%%%%%%%%%%%%%%%%%%%%%%%%%%%%
%%%%%%%%%%%%%%%%%%%%%%%%%%%%%%%%%%%%%%%%%%%%%%%%%%%
\begin{figure}[t!]
    \centering
    \includegraphics[scale=.1]{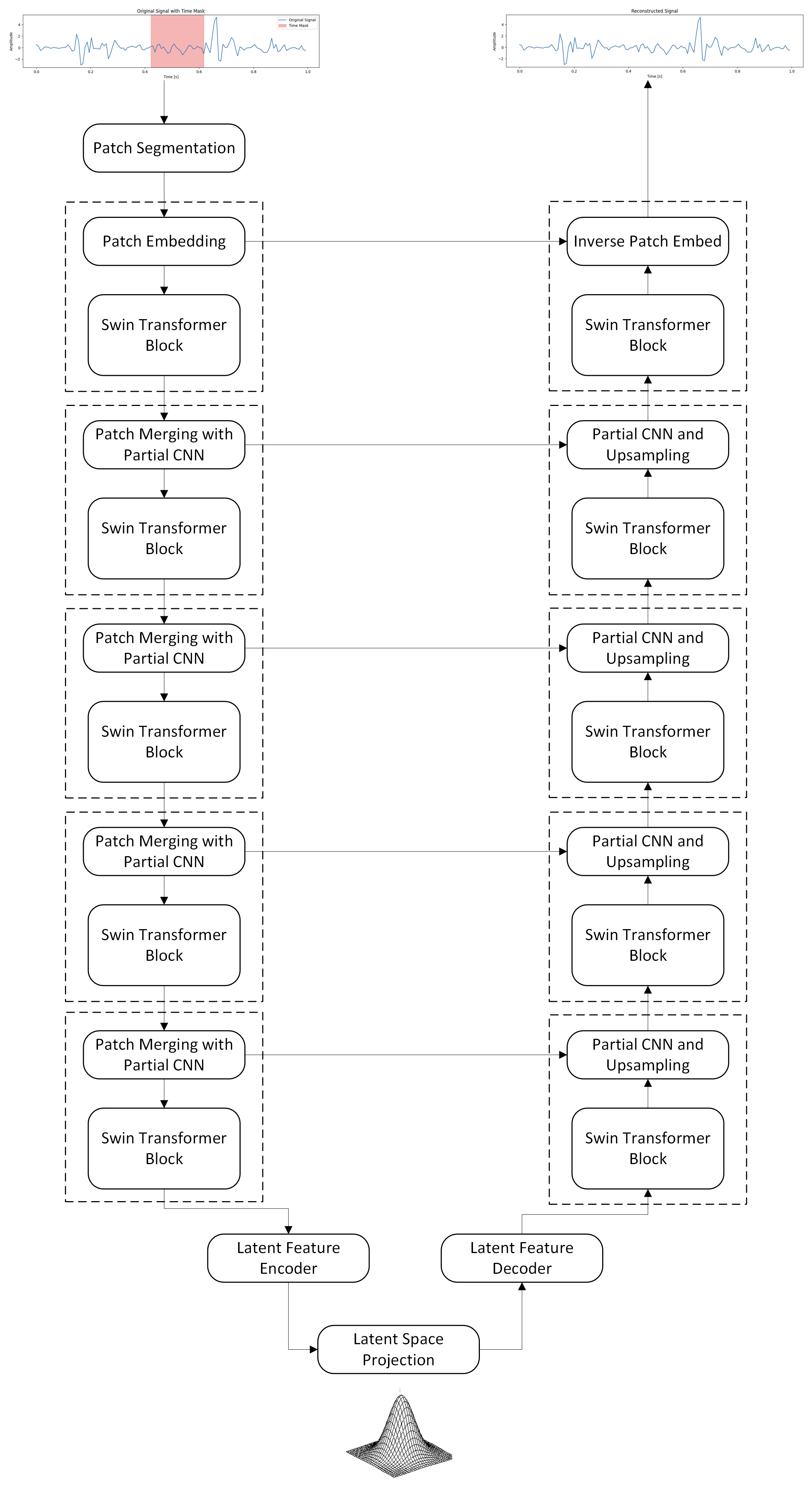}
    \caption{\footnotesize Encoder and decoder architecture of the $\KMM$ framework.}
    \label{fig:label}
\end{figure}
%%%%%%%%%%%%%%%%%%%%%%%%%%%%%%%%%%%%%%%%%%%%%%%%%%%
The overall architecture of the proposed $\KMM$ framework is shown in Fig.~\ref{fig:KMM}, with Swin-Unet~\cite{Cao2022} building blocks, which consists of an encoder, bottleneck, decoder, and skip connections. The core of the $\KMM$ architecture is built upon a three-path encoder-decoder structure, where the input data is processed through three distinct encoder pathways, each capturing different aspects of the underlying HD-sEMG signal. The outputs from these encoders are then combined and passed through a shared decoder to generate the final output. Such a multi-path approach enables the $\KMM$ to learn diverse feature representations, enhancing its ability to handle complex inputs. In each path, the HD-sEMG signals are divided into non-overlapping patches with a patch size of $4 \times 4$ and processed through several Swin Transformer blocks and patch merging layers to generate the bottleneck representation. In the first three initial layers, the extracted features are passed through skip connections to the decoder. The decoder comprises of a Swin Transformer blocks and patch expanding layers, which include upsampling layers and a fusion layer. The fusion layer utilizes several attention mechanisms and feature alignment techniques to ensure that the most important features from each domain are effectively retained and combined.

%============================================================
\subsection{Masking Strategies}
%============================================================

In the $\KMM$, we utilize a variety of masking techniques to augment the input data. These masking strategies are designed to mask portions of the input in different ways, forcing the model to learn robust representations that can handle incomplete data. Four distinct masking approaches are applied to each input, creating variations that challenge the model to infer missing information across time, sensors, and scales. The data masked in different ways is then passed through the model's encoder-decoder architecture, allowing the model to learn to reconstruct the missing information. These augmented inputs are then used in a self-supervised pre-training process to improve the model’s generalization capabilities. The following masking techniques are applied:
\begin{itemize}
\item \textit{\textbf{Random Block Masking:}} In this method, random contiguous blocks of the signal are masked. This technique helps the model learn to infer the missing blocks based on the context provided by the unmasked parts. 
\item \textit{\textbf{Temporal Masking:}} Temporal masking is applied by randomly selecting time intervals and masking them across all channels. This forces the model to understand the temporal relationships between the data points and predict the missing portions using the available temporal context.
\item \textit{\textbf{Sensor-Wise Random Masking:}} Here, random sensors (or channels) are masked entirely, simulating the scenario where certain sensors fail or do not capture data. The model must rely on the remaining sensors to reconstruct the missing sensor data, improving its robustness to missing or corrupted sensor inputs.
\item \textit{\textbf{Multi-Scale Masking:}} Multi-scale masking is designed to occlude parts of the signal at different granularities. Some masks may occlude large sections of the signal, while others mask smaller regions, forcing the model to be flexible in reconstructing both fine-grained and coarse-grained missing information.
\end{itemize}
 
%============================================================
\subsection{$\KMM$'s Encoder Architecture}
%============================================================

Separate encoder is employed for each path, allowing $\KMM$ to learn distinct yet complementary representations of the input data. In each path, the data is normalized using $Z$-score normalization, computed along the last dimension and applied separately for each sensor. More specifically, the following three distinct HD-sEMG signal variants are used each capturing a different aspect of the signal: (i) First encoder captures time-domain features from the original data; (ii) The second encoder extracts frequency-domain features from the FFT-transformed data, and; (iii) The third encoder focuses on magnitude features from the absolute values of the original data. 
%The processed HD-sEMG signals are then passed through separate encoders, with each path focusing on a different representation of the input. 

In each encoder, the data goes through multiple blocks of patch merging and transformer blocks. We employ Partial Convolution in the patch merging blocks to effectively handle masked or incomplete data, reducing the data size by half. Unlike standard convolution, which assumes that all input samples contribute equally to the output, Partial Convolution performs the operation only on the unmasked regions of the input. Its key feature is its ability to dynamically adjust for the number of valid samples under the Convolutional kernel, making it robust to missing data. A critical aspect of Partial Convolution is the propagation of the mask throughout the network. After each Partial Convolution layer, the mask is updated based on the regions where the convolution was successfully applied. This updated mask is passed to subsequent layers, allowing the network to gradually recover the missing regions as it learns to infer the missing information from the valid pixels.
The representations learned from the three paths are then integrated using two Sigmoid-Gated Cross-Modality Attention (SG-CMA) mechanisms. The first SG-CMA aligns the features between the original data and the FFT-transformed data, while the second SG-CMA aligns the original data with the absolute value data. In each SG-CMA, a sigmoid-gated fusion mechanism is applied to selectively attend to the most relevant features from each modality, enhancing the alignment and reducing the impact of less informative components. This ensures that features learned from the different transformations are not only aligned but also adaptively fused into two unified latent representations. The integration of sigmoid-gated attention allows the model to dynamically control the contributions from each modality, leveraging the strengths of each transformation (time-domain, frequency-domain, and magnitude-based features) to improve gesture recognition performance.

%============================================================
\subsection{$\KMM$'s Decoder Architecture}
%============================================================
%%%%%%%%%%%%%%%%%%%%%%%%%%%%%%%%%%%%%%%%%%%%%%%%%%%%
\begin{figure}[t!]
    \centering
    \includegraphics[width=0.45\textwidth]{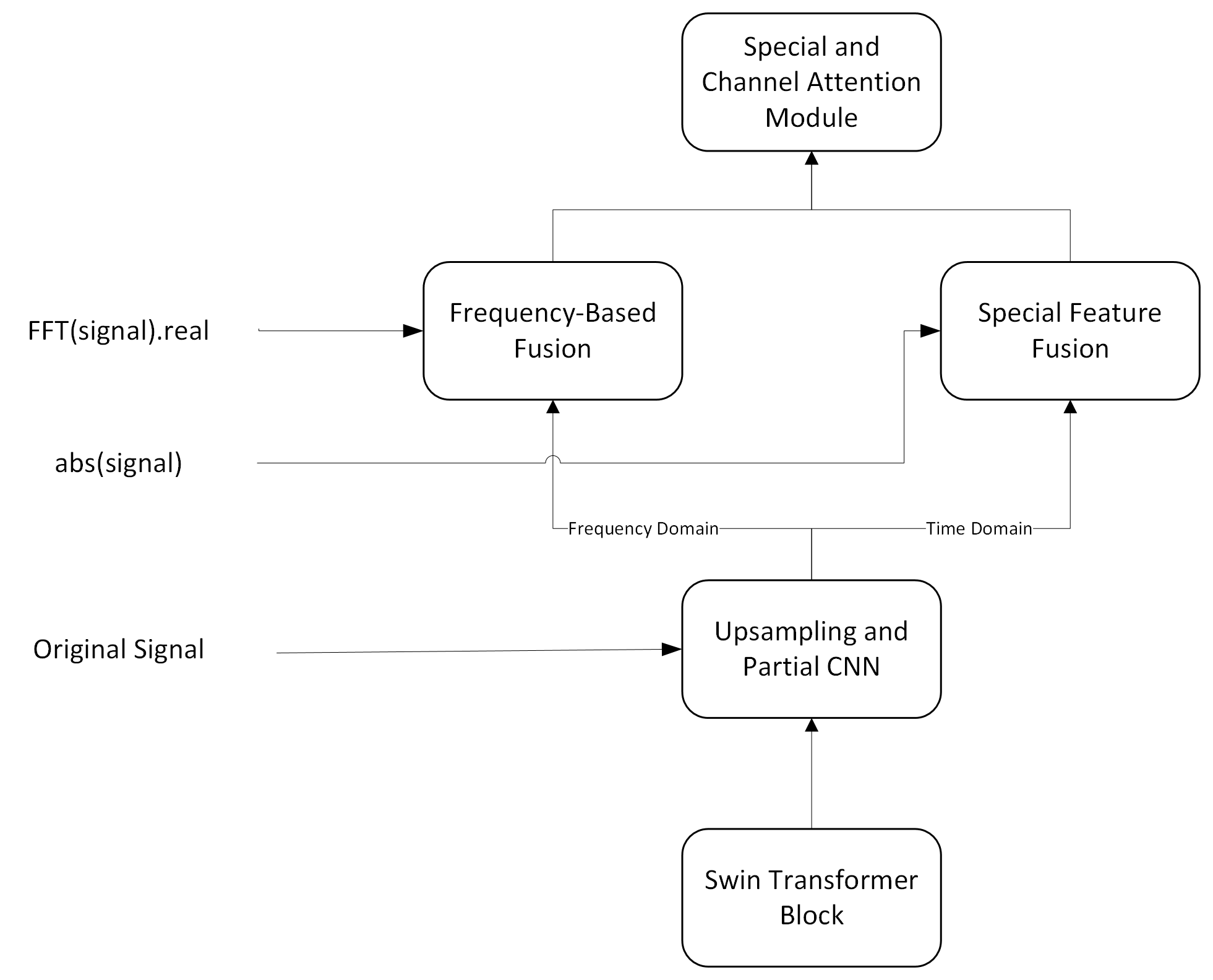}
    \caption{\footnotesize Fusion process for the first 3 Layers}
    \label{fig:fig3}
\end{figure}
%%%%%%%%%%%%%%%%%%%%%%%%%%%%%%%%%%%%%%%%%%%%%%%%%%%%

The decoder is designed to combine the multi-modal information learned by the three encoders and reconstruct the missing portions of the input. It integrates the time-domain, magnitude-based, and frequency-domain features to generate a comprehensive reconstruction of the original input, filling in the masked regions with plausible values.
The decoder consists of multiple blocks of Patch Expand layers and Transformer blocks. The Patch Expand function is a critical component of the Swin U-Net architecture, responsible for upsampling and expanding feature maps during the decoding stage. This function plays a key role in reconstructing high-resolution output from the lower-resolution features generated during the encoding stage. Patch Expand not only restores spatial resolution but also integrates multi-scale information by concatenating features from both the encoder and decoder, utilizing Partial Convolution to handle masked data.
The decoder also employs a multi-modal feature fusion strategy, which integrates information from both the spatial domain (time-domain) and the frequency domain. The fusion process utilizes several attention mechanisms and feature alignment techniques, as shown in Fig.~\ref{fig:fig3}, to ensure that the most important features from each domain are effectively combined.
The fusion process consists of frequency-based fusion and channel and spatial attention mechanisms, which together create a powerful method for combining complementary features:
\begin{itemize}
\item \textbf{\textit{Frequency-Based Fusion}} processes the input data in the frequency domain by applying convolution to extract feature maps. These maps are then transformed into the frequency domain using the Fast Fourier Transform (FFT). The resulting complex-valued frequency representation is separated into two key components: amplitude (the magnitude of the frequency components) and phase (the angular information of the frequency components). Amplitude and phase are fused using Partial Convolution, where corresponding components from two inputs (e.g., spatial and frequency representations) are concatenated and passed through convolutional layers to reconstruct the complex frequency representation, which is then transformed back into the time domain via the inverse FFT.
\item \textbf{\textit{Channel Attention Module}} applies global average pooling to the fused feature maps, followed by convolutional layers that reduce and expand the feature channels. This process generates attention weights that are applied independently to the spatial and frequency domain feature maps, allowing the network to focus on the most important features.
\item \textbf{\textit{Spatial Attention}} focuses on spatial dependencies by concatenating the spatial and frequency feature maps along the channel dimension and applying convolution. This spatial attention map highlights critical regions in both the spatial and frequency domains.
The final fusion process combines the outputs from the frequency-based and attention-based mechanisms, ensuring that the fused representation incorporates the most informative features from both domains for reconstruction and downstream tasks.
\item \textbf{\textit{Classification Head}} in our model is designed to combine features from two output embedding and produce a final classification decision. Each of the two inputs represents different feature sets, and they are concatenated to form a unified input vector. This combined vector is then passed through a series of fully connected layers.
The first layer reduces the dimensionality of the input to a more manageable size, allowing for efficient learning while maintaining relevant features. A GELU (Gaussian Error Linear Unit) activation function is applied to introduce non-linearity, which helps in capturing complex patterns in the data. Following the activation, a dropout layer is applied to prevent overfitting by randomly dropping a fraction of the neurons during training. Finally, the output is passed through a second linear layer, which maps the feature representation to the number of output classes, thus generating the final classification predictions.
\end{itemize}

%============================================================
\subsection{$\KMM$'s Training Mechanism}
%============================================================
%%%%%%%%%%%%%%%%%%%%%%%%%%%%%%%%%%%%%%%%%%%%%%%%%%%
\begin{table}[t!]
    \centering
   \caption{\footnotesize Comparision results based on the Dba dataset.  \label{tab:dba}}
    \begin{tabular}{|c|c|}
        \hline
         Model & Accuracy \\
        \hline
        TD ~\cite{Englehart-2003} & $0.920 \pm 0.050$\\
        \hline
        ETD ~\cite{Khushaba-2012} & $0.944 \pm 0.036$\\
        \hline
        NinaPro ~\cite{Atzori-2014} & $0.909 \pm 0.057$\\
        \hline
        SampEn ~\cite{Phinyomark-2013} & $0.950 \pm 0.043$\\
        \hline
        TVGGNet ~\cite{Pereira-2024-62} & $0.873 \pm 0.055$\\
        \hline
        S-ConvNet ~\cite{Islam-2020} & 0.879 \\
        \hline
        W. Geng et al. ~\cite{Geng-2016} & 0.869 \\
        \hline
        All-ConvNet ~\cite{Islam:2024} & 0.867 \\
        \hline
        The Proposed $\KMM$ & $0.973 \pm 0.0008$\\
        \hline
    \end{tabular}
\end{table}
%%%%%%%%%%%%%%%%%%%%%%%%%%%%%%%%%%%%%%%%%%%%%%%%%%%

\noindent
Training of the $\KMM$ consists of the following three-stage process:
\begin{itemize}
\item \textbf{\textit{Stage 1:}} Pre-train the MAE using the original HD-sEMG signal with a Variational Autoencoders (VAE) loss. By initially training with only the original data, the model learns the fundamental structure and key features of the raw HD-sEMG  data without being influenced by the additional complexities introduced by other  transformations. The VAE loss encourages the latent representation to follow a Gaussian distribution, organizing the latent space and making it easier to sample from. This ensures that the model develops a meaningful and smooth latent representation of the original data.
\item \textbf{\textit{Stage 2:}} Re-train the MAE by incorporating the other two encoders and introducing Cross-Modality Attention along with Contrastive loss. The Cross-Modality Attention allows the representations of the new paths to attend to the original latent features, improving alignment between the modalities. Contrastive loss and the Aggregation Mechanism~\cite{Dong2024} are also employed to encourage well-structured latent spaces, ensuring that different modalities form useful relationships for both reconstruction and downstream classification. Contrastive learning encourages the alignment of latent representations across the different paths, ensuring that similar data are close in the latent space and dissimilar data are separated. The aggregation mechanism integrates the latent representations from the three paths into a unified, meaningful representation by utilizing the similarity information derived from contrastive learning. This process improves the model's performance by leveraging complementary features from time-domain, frequency-domain, and magnitude-based representations.
\item \textbf{\textit{Stage 3:}} Finally, we fine-tune the encoder part of the network for classification.
\end{itemize}

%OOOOOOOOOOOOOOOOOOOOOOOOOOOOOOOOOOOOOOOOOOOOOOOOOOOOOOO
\section{Experimental Results} \label{sec:results}
%OOOOOOOOOOOOOOOOOOOOOOOOOOOOOOOOOOOOOOOOOOOOOOOOOOOOOOO

%============================================================
\subsection{Dataset} 
%============================================================

In this study, we used the raw HD-sEMG data from the CapgMyo dataset~\cite{CapgMyo}, which provides a rich source of EMG signals recorded through a differential, silver, wet electrode array. This array enables the simultaneous recording of data from $128$ channels, with each channel operating at a sampling rate of $1,000$ Hz. The CapgMyo dataset contains data across three sub-databases (DB-a, DB-b, and DB-c), each focused on different gesture types and recording sessions as outlined below:
\begin{itemize}
\item \textit{\textbf{DB-a:}} Contains data from $23$ subjects, each performing and holding $8$ different isotonic and isometric hand gestures for $3$ to $10$ seconds. These gestures simulate common hand movements such as thumb extension, finger flexion, and abduction.
\item \textit{\textbf{DB-b:}} Includes feature recordings from $10$ of the $23$ subjects, with each subject contributing data from two separate recording sessions (DB-b Session 1 and DB-b Session 2), recorded in a interval greater than seven days. This sub-database also includes the same $8$ hand gestures as in DB-a, with each gesture performed for approximately $3$ seconds. Inter-session variability is introduced by differences in electrode positioning at subsequent sessions.
\item \textit{\textbf{DB-c:}} Expands on the previous sub-databases by including data from $10$ of the $23$ subjects performing $12$ distinct hand gestures. Each gesture in DB-c involves more detailed finger movements, which held for approximately $3$ seconds as in DB-b.
\end{itemize}

%============================================================
\subsection{Data Pre-processing}
%============================================================

To prepare the dataset for our experiments, power-line interference was filtered out using a second-order Butterworth filter with a band-stop range between $45$ and $55$ Hz. A window size of $128$ ms with a step size of $28$ ms was used, resulting in each data segment being converted into $32$ overlapping windows of $128$ ms each. All data from the $128$ sensors were then combined to create an input of size $128 \times 128$. Finally, as the range of EMG value is between -$2.5$ mV to $2.5$ mV, normalization is applied by setting the unit to $2.5$ mV to guarantee the value remains mainly between $-1$ to $1$. 

%============================================================
\subsection{Experimental Setup}
%============================================================

Two primary experiments were conducted using different subsets of the CapgMyo dataset:
\textit{\textbf{(i) Experiment 1}} used data from sub-database DB-a, consisting of 18 subjects. Each subject performed 10 repetitions of the 8 hand gestures. Half of these repetitions were assigned to the training set, and the remaining half were used for testing.
\textit{\textbf{(ii) Experiment 2}}  utilized data from DB-b, focusing on inter-session gesture recognition. For this experiment, data from two sessions of 10 subjects were used, with one session used for training and the other for testing. Due to data corruption in the last subject’s recordings, this experiment was carried forward with the first 9 subjects~\cite{Pereira-2024-62}. 

%============================================================
\subsection{Results} 
%============================================================
%%%%%%%%%%%%%%%%%%%%%%%%%%%%%%%%%%%%%%%%%%%%%%%%%%%
\begin{table}[t!]
    \centering
     \caption{\footnotesize Comparision results based on the Dbb dataset.  \label{tab:dbb}}
    \begin{tabular}{|c|c|}
        \hline
         Model & Accuracy \\
        \hline
        TD~\cite{Englehart-2003} & $0.416 \pm 0.198$ \\
        \hline
        ETD~\cite{Khushaba-2012} & $0.437 \pm 0.209$\\
        \hline
        NinaPro~\cite{Atzori-2014} & $0.425 \pm 0.170$\\
        \hline
        SampEn~\cite{Phinyomark-2013} & $0.448 \pm 0.217$\\
        \hline
        TVGGNet~\cite{Pereira-2024-62} & $0.420 \pm 0.183$\\
        \hline
        The Proposed $\KMM$& $0.562 \pm 0.007$\\
        \hline
    \end{tabular}
\end{table}
%%%%%%%%%%%%%%%%%%%%%%%%%%%%%%%%%%%%%%%%%%%%%%%%%%%

As shown in Tables~\ref{tab:dba} and~\ref{tab:dbb}, the proposed $\KMM$ achieved acuracy of $0.973$ for DBA and $0.562$ for DBB without any domain adaptation. These are superior to the best results reported in state-of-the-art including~\cite{Islam:2024, Pereira-2024-62}. Please note that in Table~\ref{tab:dbb}, for fair comparison, we only report the available results in the literature using the same data split. For completeness, we have also included comparisons with traditional state-of-the-art solutions, i.e., Hudgin’s TD features~\cite{Englehart-2003};
%the enhanced TD features (ETD) from~\cite{Khushaba-2012}, 
features with best performance from NinaPro dataset publication~\cite{Atzori-2014}, and; the SampEn features~\cite{Phinyomark-2013} shown to provide most stable performance across sessions. In all these three cases, classification was carried out via LDA.

\begin{figure}[t!]
    \centering
    \subfigure[DBA Subset]{
        \includegraphics[width=0.48\textwidth]{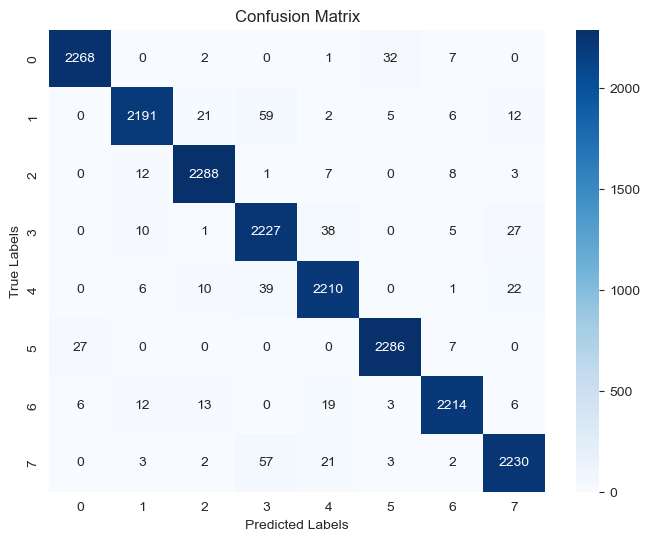} 
    }
    \hspace{0.02\textwidth}
    \subfigure[DBB Subset]{
        \includegraphics[width=0.48\textwidth]{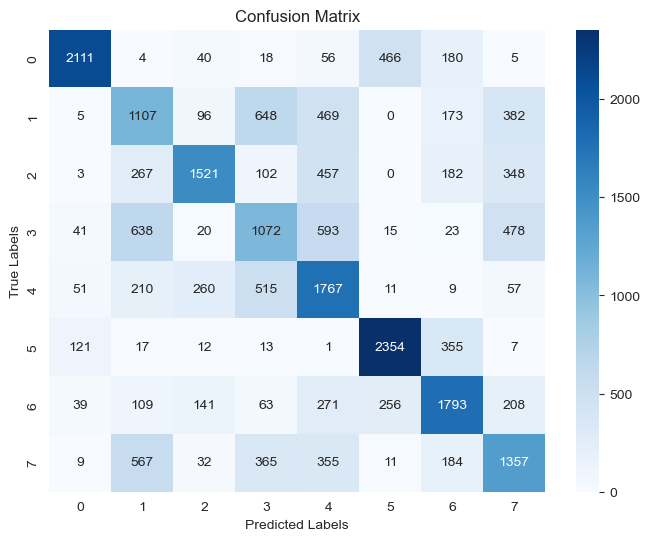}
    }
    \vspace{-.1in}
    \caption{\footnotesize Confusion matrices showing the classification performance across 8 labels. (a) The DBA subset. (b) The DBB subset.}
    \label{fig:confusion_matrices}
\end{figure}

To further analyze the model's classification performance on both subsets, we calculated the confusion matrices for each case. The confusion matrices, as illustrated in Figures 4.a and 4.b, show the distribution of predicted versus true labels for each of the 8 classes. Figure 4.a shows that the model achieves high precision and recall for most classes in the DBA subset, indicating strong performance. In contrast, Figure 4.b demonstrates lower precision and recall across several classes in the DBB subset, highlighting the impact of distributional differences between the two subsets.

We also tested our model on the DB-a subset using a segmentation approach where each sample was divided into 32 frames of 128 ms each. Using a majority voting strategy, the MAST model achieved an overall accuracy of $99.3$, outperforming state-of-the-art models such as GengNet \cite{Geng-2016, Wei-2017, Hu-2018, Du-2017, Islam-2023, Pereira-2024} and S-ConvNet \cite{Islam-2020}. The proposed approach demonstrated superior performance both in terms of per-frame accuracy and majority voting accuracy.

For the DB-b subset, which often suffers from significant performance degradation due to electrode shifts between sessions, our baseline MAE-based model achieved an accuracy of 56.2\% under inter-session conditions, outperforming existing non-domain adaptation-based methods such as \cite{Du-2017}, who reported 41.2\% per frame accuracy, and \cite{Ketyko-2019}, who achieved 53\% using a two-stage recurrent neural network (2SRNN) approach. To further enhance the inter-session performance, we incorporated a small amount of session 2 data (10\%) into the training set as a form of domain adaptation. This strategy effectively mitigated the effects of electrode shifts, improving our model's accuracy to 78.2\%, surpassing the results reported by \cite{Du-2017} and \cite{Pereira-2024}, who achieved accuracies of 67.97\% and 75.91\%, respectively, when using 20\% of session 2 data for adaptation.

Our MAE pretraining approach also demonstrated a notable improvement in generalization ability. By pretraining the MAE with a mask ratio of 50\%, we observed a 4\% improvement in accuracy on the DB-a subset and a 6\% improvement on DB-b. These findings highlight the benefits of self-supervised learning strategies like MAE pretraining, which significantly enhance the model's ability to generalize across different conditions, particularly in inter-session scenarios. Overall, the proposed MAST framework consistently outperformed state-of-the-art methods across both subsets of the CapgMyo dataset, demonstrating its robustness and superior generalization capability.

%OOOOOOOOOOOOOOOOOOOOOOOOOOOOOOOOOOOOOOOOOOOOOOOOOOOOOOO
\vspace{-.1in}
\section{Conclusions} \label{sec:conc}
\vspace{-.05in}
%OOOOOOOOOOOOOOOOOOOOOOOOOOOOOOOOOOOOOOOOOOOOOOOOOOOOOOO
In conclusion, the proposed $\KMM$ framework targets addressing the challenge of performance degradation in HD-sEMG-based gesture recognition by changing recording conditions. By employing a combination of four masking strategies and utilizing a multi-path Swin-Unet architecture, the $\KMM$ framework effectively captures diverse features of the HD-sEMG signal. The self-supervised pre-training approach further enhances the model's generalization capabilities, leading to a considerable improvement in intersession performance. 

\bibliographystyle{IEEEbib}
%\bibliography{refs}

\begin{thebibliography}{10}
\footnotesize

\bibitem{Bi-2019-2}
L. Bi, and C. Guan,
\newblock ``{A Review on EMG-based Motor Intention Prediction of Continuous Human Upper Limb Motion for Human-Robot Collaboration,}''
\newblock {\em Biomed. Signal Process. \& Control}, 51, pp.113-127.~2019.

\bibitem{Farina-2014-4}
D. Farina, \textit{et al.},
\newblock ``{The Extraction of Neural Information from the surface EMG for the Control of Upper-Limb Prostheses: Emerging Avenues and Challenges,}''
\newblock {\em IEEE Trans. Neural Syst. Rehabil. Eng.}, vo. 22, no. 4, pp.797-809. 2014.

\bibitem{Islam:2024}
R. Islam, D. Massicotte, P. Massicotte and W.P. Zhu, 
\newblock ``Surface EMG-Based Intersession/Intersubject Gesture Recognition by Leveraging Lightweight All-ConvNet and Transfer Learning," 
\newblock {\em IEEE Trans. Instrum. Meas.}, vol. 73, pp. 1-16, 2024.

\bibitem{Farokh:2024}
Q. Hu, \textit{et al.}, 
\newblock ``ViT-MDHGR: Cross-Day Reliability and Agility in Dynamic Hand Gesture Prediction via HD-sEMG Signal Decoding," 
\newblock {\em IEEE J. Sele. Topics Signal Process.}, vol. 18, no. 3, pp. 419-430, 2024.

\bibitem{Kia:2024}
S.A. Stuttaford, M. Dyson, K. Nazarpour, S.S.G. Dupan, 
\newblock ``Reducing Motor Variability Enhances Myoelectric Control Robustness Across Untrained Limb Positions," 
\newblock {\em IEEE Trans. Neural Syst. Rehabil. Eng.}, vol. 32, pp. 23-32, 2024.

\bibitem{Li-2020-6}
Z. Li, \textit{et al.},
\newblock ``{Electrode Shifts Estimation and Adaptive Correction for Improving Robustness of sEMG-based Recognition,}''
\newblock {\em IEEE J. Biomed. \& Health Informat.}, vol. 25, no. 4, pp.1101-1110, 2020.

\bibitem{Ding-2019-7}
Q. Ding,  \textit{et al.},
\newblock ``{Adaptive Hybrid Classifier for Myoelectric Pattern Recognition against the Interferences of Outlier Motion, Muscle Fatigue, and Electrode Doffing,}''
\newblock {\em IEEE Trans. Neural Syst. Rehabil. Eng.}, vol. 27, no. 5, pp.1071-1080, 2019.

\bibitem{Hargrove-2006-8}
L. Hargrove, K. Englehart, and B. Hudgins,
\newblock ``{The effect of Electrode Displacements on Pattern Recognition based Myoelectric Control,}''
\newblock {\em Int. Con. of IEEE Eng. in Medicine \& Biology Soc.}, 2016.

\bibitem{Zhu-2016-20}
X. Zhu,  \textit{et al.},
\newblock ``{Cascaded Adaptation Framework for Fast Calibration of Myoelectric Control,}''
\newblock {\em  IEEE Trans. Neural Syst. Rehabil. Eng.}, vol. 25, no. 3, pp.254-264, 2016.

\bibitem{Zhang-2019-13}
X. Zhang, \textit{et al.},
\newblock ``{Adaptive Calibration of Electrode Array Shifts enables Robust Myoelectric Control,}''
\newblock {\em IEEE Trans. Biomed. Eng.}, vol. 67, no. 7, pp.1947-1957, 2019.

\bibitem{Li-2020-25}
K. Li, \textit{et al.},
\newblock ``{A Review of the Key Technologies for sEMG-based Human-Robot Interaction Systems,}''
\newblock {\em Biomed. Signal Process. \& Control}, vol. 62, p.102074, 2020.

\bibitem{Stango-2014-32}
A. Stango, F. Negro, and D. Farina, 
\newblock ``{Spatial Correlation of High Density EMG Signals Provides Features Robust to Electrode Number and Shift in Pattern Recognition for Myocontrol,}''
\newblock {\em IEEE Trans. Neural Syst. Rehabil. Eng.}, vol. 23, no. 2, pp.189-198, 2014.

\bibitem{ElahehTNSRE:2021}
E. Rahimian, S. Zabihi, A. Asif, D. Farina, S. F. Atashzar and A. Mohammadi, 
\newblock ``FS-HGR: Few-Shot Learning for Hand Gesture Recognition via Electromyography," 
\newblock {\em IEEE Trans. Neural Syst. Rehabil. Eng.}, vol. 29, pp. 1004-1015, 2021.

\bibitem{Pereira-2024-62}
J. Pereira, D. Halatsis, B. Hodossy, and D. Farina, 
\newblock ``Tackling Electrode Shift in Gesture Recognition with HD-EMG Electrode Subsets,''
\newblock {\em IEEE Int. Con. Acoustics, Speech \& Signal Process. (ICASSP)},  2024.

\bibitem{Elaheh:G2019}
E. Rahimian, S. Zabihi, S. F. Atashzar, A. Asif and A. Mohammadi, 
\newblock ``sEMG-based Hand Gesture Recognition Via Dilated Convolutional Neural Networks," \newblock {\em IEEE Global Conf. Signal \& Inf. Process.}, 2019.

%%%%

\bibitem{Rami:2024}
F. Kulwa, \textit{et al.}, 
\newblock ``A Multidataset Characterization of Window-Based Hyperparameters for Deep CNN-Driven sEMG Pattern Recognition," 
\newblock {\em IEEE Trans. Human-Mach. Syst.}, vol. 54, no. 1, pp. 131-142, 2024

\bibitem{Zhang-2020-54}
Y. Zhang, \textit{et al.}, 
\newblock ``{Learning Effective Spatial-temporal features for sEMG armband-based gesture recognition},''
\newblock {\em IEEE Internet of Things J.}, vol. 7, no. 8, pp. 6979-6992, 2020.

\bibitem{Karnam-2022-55}
N.K. Karnam, S.R. Dubey, A.C. Turlapaty, B. Gokaraju,
\newblock ``{EMGHandNet: A hybrid CNN and Bi-LSTM Architecture for Hand Activity Classification using surface EMG Signals},''
\newblock {\em Biocybernetics \& Biomed. Eng.}, vol. 42, no. 1, pp. 325-340, 2022.

\bibitem{Kia:2020}
M. Jabbari, R.N. Khushaba, K. Nazarpour, 
\newblock ``EMG-Based Hand Gesture Classification with Long Short-Term Memory Deep Recurrent Neural Networks," 
\newblock {\em Int. Con. of IEEE Eng. in Medicine \& Biology Soc.}, pp. 3302-3305, 2020.

\bibitem{Sun-2022-56}
B. Sun, \textit{et al.}, 
\newblock ``A Multiscale Feature Extraction network based on Channel-Spatial Attention for Electromyographic Signal Classification,''
\newblock {\em IEEE Trans. Cognitive \& Developmental Syst.}, vol. 15, no.2 , pp. 591-601, 2022.

\bibitem{Qureshi-2023-57}
M.F. Qureshi, \textit{et al.},
\newblock ``{E2cnn: An efficient concatenated cnn for classification of surface emg extracted from upper limb},''
\newblock {\em IEEE Sensors Journal}, 23(8), pp.8989-8996. 2023.


\bibitem{Han-2021-60}
L. Han, Y. Zou, L. Cheng, 
\newblock ``{A Convolutional Neural Network with Multi-scale Kernel and Feature Fusion for sEMG-based Gesture Recognition},''
\newblock {\em  IEEE Int. Con. Robotics \& Biomimetics (ROBIO)}, pp. 774-779. 2021.

\bibitem{Shen-2020-61}
S. Shen, \textit{et al.},
\newblock ``Gesture Recognition through sEMG with Wearable Device based on Deep Learning,''
\newblock {\em Mobile Networks and Applications}, 25, pp. 2447-2458, 2020.

\bibitem{Montazerin-2023-14}
M. Montazerin, \textit{et al.},
\newblock ``{Transformer-based Hand Gesture Recognition from Instantaneous to Fused Neural Decomposition of High-Density EMG Signals,}''
\newblock {\em Scientific Reports}, vol. 13, no. 1, p.11000, 2023.

\bibitem{MansooreEMBC:2022}
M. Montazerin, \textit{et al.},
\newblock ``ViT-HGR: Vision Transformer-based Hand Gesture Recognition from High Density Surface EMG Signals," 
\newblock {\em Int. Con. of IEEE Eng. in Medicine \& Biology Soc.}, 2022, pp. 5115-5119.

\bibitem{SoheilTNSR2023}
S. Zabihi, E. Rahimian, A. Asif and A. Mohammadi, 
\newblock ``TraHGR: Transformer for Hand Gesture Recognition via Electromyography," 
\newblock {\em IEEE Trans. Neural Syst. Rehabil. Eng.}, vol. 31, pp. 4211-4224, 2023.

\bibitem{Hao-2021-58}
S. Hao, R. Wang, Y. Wang, Y. Li, 
\newblock ``A Spatial Attention based Convolutional Neural Network for Gesture Recognition with HD-sEMG Signals,''
\newblock {\em  IEEE Int. Con. E-health Net., Application \& Services}, 2021.

\bibitem{Wang-2022-59}
L. Wang, J. Fu, B. Zheng, H. Zhao, 
\newblock ``Research on sEMG-based Gesture Recognition using the Attention-based LSTM-CNN with Stationary Wavelet Packet Transform,''
\newblock {\em Int. Con. Advances in Computer Technology, Inf. Science \& Commun. (CTISC)}, pp. 1-6. 2022.

\bibitem{Wang-2023-60}
Wang, L., Li, X., Chen, Z., Sun, Z. and Xue, J., 
\newblock ``Electrode shift fast adaptive correction for improving myoelectric control interface performance,''
\newblock {\em IEEE Sensors Journal}, 2023.

\bibitem{Chamberland-2023-61}
F. Chamberland, \textit{et al.},
\newblock ``EMaGer: A Wearable Full-Circumference HD-EMG Sensor and Data Augmentation Method for Robust Hand Gesture Recognition,''
\newblock {\em Int. Con. of IEEE Eng. in Medicine \& Biology Soc.}, pp. 1-5, 2023.

%\bibitem{Chamberland-2023-63}
%F. Chamberland, \textit{et al.},
%\newblock ``Novel wearable HD-EMG sensor with shift-robust gesture recognition using deep learning,''
%\newblock {\em IEEE Transactions on Biomedical Circuits and Systems}, 2023.

%\bibitem{Wang-2024-64}
%Wang, B., Li, J., Hargrove, L. and Kamavuako, E.N., 
%\newblock ``Unravelling Influence Factors in Pattern Recognition Myoelectric Control Systems: The Impact of Limb Positions and Electrode Shifts,''
%\newblock {\em Sensors}, 24(15), p.4840., 2024.

\bibitem{He:2022}
K. He, \textit{et al.}, 
\newblock ``Masked Autoencoders are Scalable Vision Learners,''
\newblock {\em IEEE/CVF Con. Computer Vision \& Pattern Recognition (CVPR)}, pp. 16000-16009, 2022.

\bibitem{CapgMyo}
Y. Du, W. Jin, W. Wei, Y. Hu, W. Geng, 
\newblock ``Surface EMG-based Inter-Session Gesture Recognition Enhanced by Deep Domain Adaptation,''
\newblock {\em Sensors}, vol. 17, no. 3, 2017.

\bibitem{Cao2022}
H. Cao, \textit{et al.}, 
\newblock ``Swin-Unet: Unet-like Pure Transformer for Medical Image Segmentation,''
\newblock {\em European Conference on Computer Vision}, pp. 205-218, 2022.

\bibitem{Dong2024}
J. Dong, \textit{et al.}, 
\newblock ``{SimMTM: A Simple Pre-Training Framework for Masked Time-Series Modeling},''
\newblock {\em Advances in Neural Information Processing Systems}, vol. 36., 2024.

\bibitem{Englehart-2003}
K. Englehart, B. Hudgins, 
\newblock ``A Robust, Real-time Control Scheme for Multifunction Myoelectric Control,''
\newblock {\em IEEE Trans. Biomed. Eng.}, vol. 50, no. 7, pp.848-854, 2003.

\bibitem{Khushaba-2012}
Khushaba, R.N. and Kodagoda, S., 
\newblock ``{Electromyogram (EMG) feature reduction using mutual components analysis for multifunction prosthetic fingers control},''
\newblock {\em 12th International Conference on Control Automation Robotics \& Vision (ICARCV)}, pp. 1534-1539, 2012.

\bibitem{Atzori-2014}
M. Atzori, \textit{et al.}, 
\newblock ``Electromyography Data for Non-invasive Naturally-Controlled Robotic hand Prostheses,''
\newblock {\em Scientific data}, pp.1-13, 2014.

\bibitem{Phinyomark-2013}
A. Phinyomark, \textit{et al.},
\newblock ``EMG Feature Evaluation for Improving Myoelectric Pattern Recognition Robustness,''
\newblock {\em Expert Systems with Applications}, 40(12), pp.4832-4840. 2013.

\bibitem{Islam-2020}
M.R. Islam, \textit{et al.},
\newblock ``S-ConvNet: A Shallow Convolutional Neural Network Architecture for Neuromuscular Activity Recognition using Instantaneous High-Density surface EMG Images,''
\newblock {\em IEEE Int. Con. Eng. in Medicine \& Biology Soc.}, pp. 744-749, 2020.

\bibitem{Geng-2016}
W. Geng, \textit{et al.},
\newblock ``{Gesture Recognition by Instantaneous surface EMG Images},''
\newblock {\em Scientific reports}, 6(1), 2016.

\bibitem{Wei-2017}
W. T. Wei, Y. Hu, W. Jin, and Y. Du,
\newblock ``A multi-stream convolutional neural network for sEMG-based gesture recognition,''
\newblock {\em Pattern Recognition Letters}, vol. 119, pp. 131-138, 2017.

\bibitem{Hu-2018}
Y. Hu, W. T. Wei, W. Jin, and Y. Du,
\newblock ``A novel attention-based hybrid CNN-RNN architecture for sEMG-based gesture recognition,''
\newblock {\em PLoS ONE}, vol. 13, no. 10, e0206049, 2018.

\bibitem{Du-2017}
Y. Du, W. Jin, W. T. Wei, Y. Hu, and W. Geng,
\newblock ``Surface EMG based inter-session gesture recognition enhanced by deep domain adaptation,''
\newblock {\em Sensors}, vol. 17, no. 3, 2017.

\bibitem{Islam-2023}
Md R. Islam, D. Massicotte, and W. Zhu,
\newblock ``Surface EMG-Based Inter-Session Gesture Recognition,''
\newblock {\em arXiv preprint arXiv:2302.03948}, 2023.

\bibitem{Pereira-2024}
J. Pereira, M. R. Islam, W. Zhu, and D. Massicotte,
\newblock ``Tackling Electrode Shift in Gesture Recognition with HD-EMG Electrode Subsets,''
\newblock {\em Proc. IEEE International Conference on Acoustics, Speech and Signal Processing (ICASSP)}, 2024.

\bibitem{Ketyko-2019}
I. Ketykó, B. Van den Broeck, A. Defoort, M. M. Van Hulle, and J. Duysens,
\newblock ``Domain adaptation for sEMG-based gesture recognition with recurrent neural networks,''
\newblock {\em arXiv preprint arXiv:1903.08484}, 2019.



\end{thebibliography}

\end{document}